# Dielectric layer dependent surface plasmon effect of metallic nanoparticles on silicon substrate[*]

Xu Rui（徐锐）[a], Wang Xiao-Dong（王晓东）[a)**], Liu Wen（刘雯）[a], Xu Xiao-Na（徐晓娜）[a], Li Yue-Qiang(李越强)[a], Ji An（季安）[a], Yang Fu-Hua（杨富华）[a, b], Li Jin-Min（李晋闽）[c]

[a] *Engineering Research Center for Semiconductor Integrated Technology, Institute of Semiconductors, Chinese Academy of Sciences, Beijing 100083, P. R. China*

[b] *The State Key Laboratory for Superlattices and Microstructures, Institute of Semiconductors, Chinese Academy of Sciences, Beijing 100083, P. R. China*

[c] *Key Laboratory of Semiconductor Materials Science, Institute of Semiconductors, Chinese academy of sciences, Beijing 100083, China*

## Abstract

The electromagnetic interaction between Ag nanoparticles on the top of the Si substrate and the incident light has been studied by numerical simulations. It is found that the presence of dielectric layers with different thicknesses lead to varied resonance wavelength and scattering cross section, and consequently shifted photocurrent response over all wavelengths. These different behaviors are determined by whether the dielectric layer is beyond the domain where the elcetric field of metallic plasmons takes effect, combined with the effect of geometrical optics. It is revealed that for particle of a certain size, an appropriate dielectric thickness is desirable to achieve the best absorption. For a certain thickness of spacer, an appropriate granular size is also desirable. These observations have substantial applications for the optimization of surface plasmon enhanced silicon solar cells.

[*] Supported by the National Basic Research Program of China (973 Program) under the grant number 2010CB934104, 2010CB933800 and National Natural Science Foundation of China under the grant number of 60606024, 61076077

[**] Corresponding author E-mail: xdwang@semi.ac.cn

**PACC:** 7320M, 7360H (**PACS:** 52.40.Hf, 52.40.Fd)

# 1. Introduction

Photovoltaics is emerging as an important technology for the future energy production. To fully realize this potential, it is necessary to realize a high conversion efficiency for the solar cells. In recent years, surface plasmon is taken as one of the best solutions to achieve this object [1-7]. The emerging field of plasmonics has yielded methods for guiding and localizing light at the nanoscale, well below the wavelength of light in free space [8,9]. Surface plasmon resonance can be interpreted as the collection of electrons at the metallic surface, induced by the electromagnetic wave, and these collected electrons vibrate together in a particular way. Due to their evanescent nature, surface plasmons are able to give rise to peculiar interaction with light, leading to effective light localization and concentration.

The nanoparticle size, shape, material, as well as the local dielectric environment can affect the light scattering and coupling efficiency [10-12]. Noble metals such as Ag, Au, Cu are suitable to support surface plasmonic propagation and resonance [13], because they exhibit negative values of real part of the dielectric permittivity. Ag is a better choice than Au, due to its lower absorption and cost, although it must be well encapsulated to avoid oxidation effect that is not present for Au. Cu is cheaper but is quite absorbing. There has been a great deal of work about the effect of particle dimension on surface plasmon. Marrocco et al. indicated that the plasmon resonance is affected by a redshift as the nanoparticle diameter is progressively increased. Moreover, the scattering cross section increases with the particle size and the increase of the scattering cross section induces an enhanced forward scattering [14]. Small particles are more absorbing, and dynamic depolarization and radiation damping become important for the large particle [15]. Furthermore, the excitation higher-order plasmon modes must be taken into account for the large size [16-18]. Mertens et al. showed there is an optical nanosphere diameter range for luminescence quantum efficiency enhancement associated with resonant coupling to plasmon modes [18]. As for the particle shape, recent work by Ma and his associates indicated an exponential blue shift of resonance wavelength and a decay of extinction efficiency with a change in vertex angle of delta-stars. Because of the electric field coupling effect between neighboring vertices, electric field near the vertices interacts with each other when the vertices are fabricated closely, resulting in a distinct blue shift of resonant wavelength [19]. Catchpole et al. showed that the fraction of incident light coupled into the substrate reduces with the increase of the average spacing from nanoparticle to the substrate, but the effective scattering cross section increases [20, 21]. Currently, the effect of the underlying dielectric environment on surface plasmon has attracted a great deal of attention. Three different dielectrics are commonly used in photovoltaic cell fabrication: silicon dioxide ($SiO_2$), silicon nitride ($Si_3N_4$), and titanium dioxide ($TiO_2$), with refractive indices of 1.5, 2.0, and 2.5, respectively. Beck et al. indicated the surface plasmon resonance is redshifted by varying the underlying refractive index by up to 200 nm, from a wavelength of 500 nm for the particle on $SiO_2$ to 700 nm for that on $TiO_2$ [22].

Highly-accurate simulation is capable of facilitating the success in diverse application areas. It reduces reliance upon costly experimental prototypes, leading to a quicker assessment of design concepts. At present, several numerical techniques have been developed to investigate the interaction of light with particles. Discrete Dipole Approximation [DDA] methods and Finite Difference Time Domain [FDTD] can be listed as two of the most widely used approaches to the problem. FDTD is based on Yee lattices, in which the vector components of the electric field and magnetic field are spatially staggered. In the FDTD method, a leapfrog scheme is proposed for marching in time wherein the electric field and magnetic field updates are staggered so that Electric field updates are conducted midway during each time-step between successive magnetic field updates, and conversely. FDTD can model material through parameter averaging or, choosing grids to conform to the geometries of the material boundaries. The main advantages of FDTD-based techniques for solving electromagnetic problems are simplicity and the ability to handle complex geometries.

Placing nanoparticles on a substrate can affect the light coupling into substrate in three ways: 1) modifying the plasmon resonance wavelength, 2) changing the normalized scattering cross section, 3) changing the angular spectrum of the scattered light. In numerical simulations, the position of the normalized scattering cross section [$Q_{scat}$] peak indicates the plasmon resonance wavelength, and the fraction of the forward scattered light [$F_{sub}$] shows the angular spectrum of the scattered light. As we know, the absorbed light in the long wavelength range into the nanoparticles is negligibly small. So the $F_{sub}$ can be replaced by the light coupling efficiency into the Si substrate [$CPE$]. The normalized scattering cross section can be calculated as

$$Q_{scat} = \frac{8}{3} k^4 r^4 \left| \frac{\varepsilon - \varepsilon_m}{\varepsilon + 2\varepsilon_m} \right|^2 \quad (1),$$

where k is the incident light wave vector, r is the diameter of nanoparticle, $\varepsilon_m$ is the dielectric permittivity of dielectric environment, the dielectric permittivity of the nanoparticle is further described by $\varepsilon$ [23]. The particle scattering cross section is the largest when $\varepsilon = -2\varepsilon_m$.

It is well known that the effective range of surface plasmon is as short as only several or tens of nanometers. Therefore, the spacer thickness between the nanoparticles and the Si substrate is an important parameter. To our knowledge, few papers focus on the thickness of the dielectric layer between the nanoparticles and the absorption substrate. Through numerical calculation of the $Q_{scat}$ and *CPE*, we present systemically the spacer-thickness dependent surface plasmon effect of Ag nanoparticles and varying radius of nanoparticle as a means of enhancing light trapping in silicon solar cells. It is found that a suitable dielectric thickness for particle with specific size and an appropriate radius for a spacer of certain thickness are desirable to achieve the best absorption performance.

## 2. NUMERICAL ANALYSIS

### 2.1. Varying the thickness of the SiO$_2$ layer

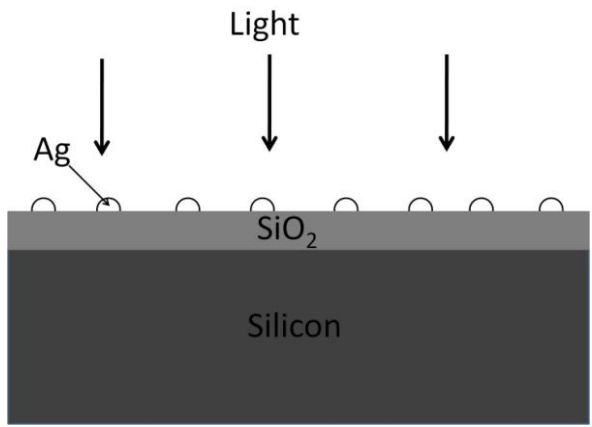

Fig. 1 The layout of the simulation design

Fig. 1 shows the layout of the simulation design. In this configuration, Ag is selected as the material to support the surface plasmon resonance, and the metallic particles are separated from a semi-infinite Si substrate by a thin SiO$_2$ layer. Numerical simulations are performed using the FDTD solutions package from Lumerical software. Nanoscale Ag semisphere arrays on a Si substrate are modeled using periodic boundary conditions and the Z dimension is truncated by the perfectly matched layers. To clarify the size effect, two sets of semispherical radii 30 nm and 70 nm are modeled with the corresponding periods of 120 and 240 nm. The mesh size around the particles is 1.5 nm and 2.5 nm, respectively. A normally incident Total-field scattered-field source [TFSF] is used to illuminate the substrate. The semisphere is completely localized in the total field. The dielectric functions are modeled using a Drude model for Ag and a Drude-Lorentz model for Si. To calculate the normalized scattering cross section the Poynting vector of the scattered field is integrated over a box surrounding the nanoparticle and then divided by the incident intensity and by the geometrical cross section. The integrated Poynting vector of the scattered field is calculated separately in the air and in the substrate. To calculate the power absorbed in the Si, a power monitor is used, located at the top surface of the Si substrate, where Si is supposed to be infinite so that the light incident into the Si substrate is completely absorbed. A normally incident plane wave with a wavelength range from 400 nm to 1100 nm illuminates the sample.

The *CPE* into the substrate is evaluated as the integral of the Poynting vector of the interface between Si substrate and the dielectric layer divided by the power of the source. Using *CPE*, integrated quantum efficiency [*IQE*] is defined as

$$IQE = \frac{\int \frac{\lambda}{hc} CPE(\lambda) I_{AM1.5}(\lambda) d\lambda}{\int \frac{\lambda}{hc} I_{AM1.5}(\lambda) d\lambda} \quad (2)$$

where *h* is Plank's constant, c is the speed of light in the free space, λ is the sunlight wavelength and $I_{AM1.5}$ is solar spectrum of AM1.5 in which sunlight radiating through an air mass is 1.5 times greater than the vertical case. In Eq. (2), numerator and denominator mean the number of photons absorbed by the Si substrate and that falling onto the Si substrate, respectively. The sun spectrum $I_{AM1.5}$ is taken from the link [24].

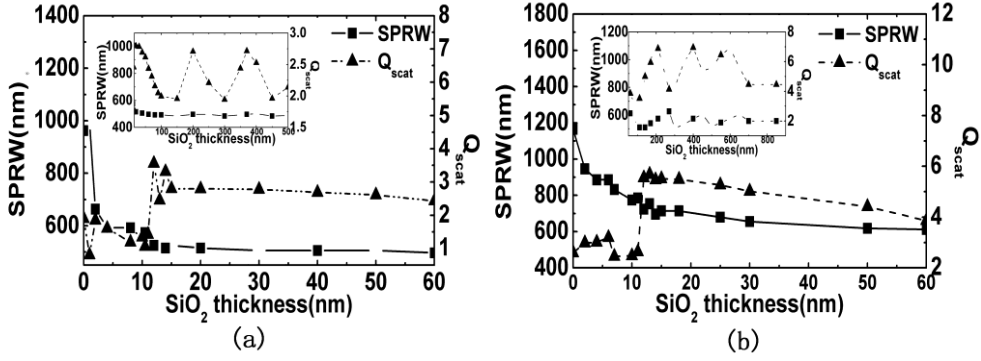

Fig. 2 The normalized scattering cross section [$Q_{scat}$] and surface plasmon resonance wavelength [$SPRW$] of semi-spherical particles vs. the thickness of $SiO_2$, for two different radii: (a) r=30 nm and (b) r=70 nm. Insets show the whole profiles.

Figure 2 illustrates the sensitivity of the normalized $Q_{scat}$ peak and the surface plasmon resonance wavelength [$SPRW$] of the nanoparticles with 30 nm and 70 nm radii to the geometry of the dielectric layer. The resonance peak blueshifts rapidly with the increase of the spacer thickness, in particular for the small particle, demonstrating the tunability of the plasmon resonance in oxide thickness. The resonance frequency can be tuned by varying the dielectric constant of the dielectric environment: a higher index leads to a redshift of the resonance [25, 26]. Clearly, increased overlap of the metallic plasmon with the high-index Si substrate causes a redshift in resonance. For Ag nanoparticle with 70 nm radius on bare Si, the resonance peak strongly redshifts beyond a wavelength of 1100 nm. The redshift becomes very slow as the $SiO_2$ layer thickness is larger than a certain value, e.g. 15 nm for the two cases, which can be ascribed to the localization of surface plasmon. For nanospheres in a static electric field, the electrical field outside the sphere [$E_{out}$] decreases rapidly, due to the rapid scaling of $E_{out} \propto 1/r^{-3}$ [23]. Therefore, the effect of Si substrate on the metallic plasmons is almost eliminated as the $SiO_2$ is thick enough. As the thickness of $SiO_2$ is beyond 100 nm, normally the same order as the visible light, where the $SiO_2$ layer can be used as an antireflection film, the $Q_{scat}$ and SPRW can be characterized by oscillations under the influence of the geometrical optics.

As can be seen in figure 2, the value of $Q_{scat}$ changes significantly and some oscillations are observed as the thickness of the dielectric layer thickness is nearby a value (about 10-15 nm), indicating the metallic plasmon is determined by both the $SiO_2$ and the Si substrate. As the oxide is thick enough, the $Q_{scat}$ remained approximately constant. In other word, the effect of the Si substrate on plasmon is quite small. The metallic plasmon depends on the oxide thickness, suggesting that modification of the spacer thickness is an effective method of prompting the optical absorption.

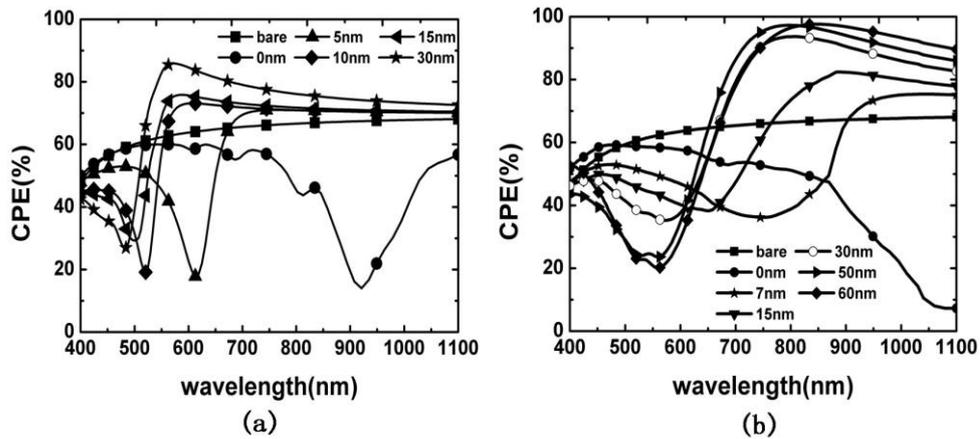

Fig. 3 The light coupling efficiency [CPE] into the substrate coated by $SiO_2$ layer with different thicknesses as a function of wavelength for two configurations: (a) radius=30 nm and (b) radius=70 nm. The numbers inside the pictures indicate the thicknesses of $SiO_2$ layers.

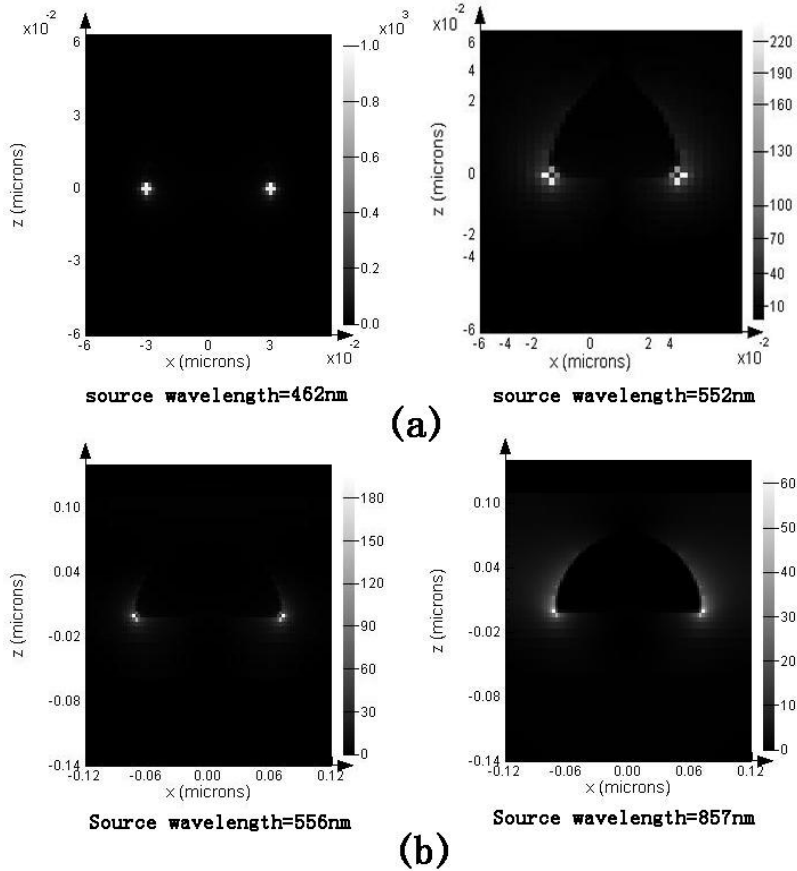

Fig.4 The electric intensity at vertical cross-sections of semispheres with the radii of (a) 30 nm and (b) 70 nm, illuminated by the incident light with different wavelengths.

To further illustrate this phenomenon, figure 3 shows the fraction of light directed into the substrate coated by oxide with different thicknesses. As can be seen in figure 3, a large fraction of the light is coupled into the substrate, in agreement with figure 1 where an effective cross section is well in excess of the geometrical area [$Q_{scat}$>1]. A maximum and minimum in the absorption spectrum is observed in figure 3, which localizes both sides of the resonance wavelength. Compared to the bare substrate, enhancement is seen only for wavelengths above certain values, such as 520 nm for 30 nm of oxide in figure 3(a). Large reduction in absorption, as much as 45%, is observed for the 30 nm radius particle underlain by 5 nm thick spacer at a wavelength of approximate 600 nm. The reduction in absorption around resonance is attributed to two mechanisms. First, interference between light that transmits at the Si interface and the scattered light by the nanoparticles. For wavelengths below resonance, the scattered light is out of phase with the incident light, leading to destructive interference [27]. Second, high electric intensity locates at the corner and the light is not efficiently conveyed into the substrate in the visible range, as can be seen in figure 4. Figure 4 gives the electric intensity distribution at vertical cross sections of Ag nanoparticles with different sizes. Left pictures are for the absorption minimum, while right pictures are for the absorption maximum. The absorption maximum can be attributed to the effectively unfolding of light in the substrate and the constructive interference between the scattered light and the incident light.

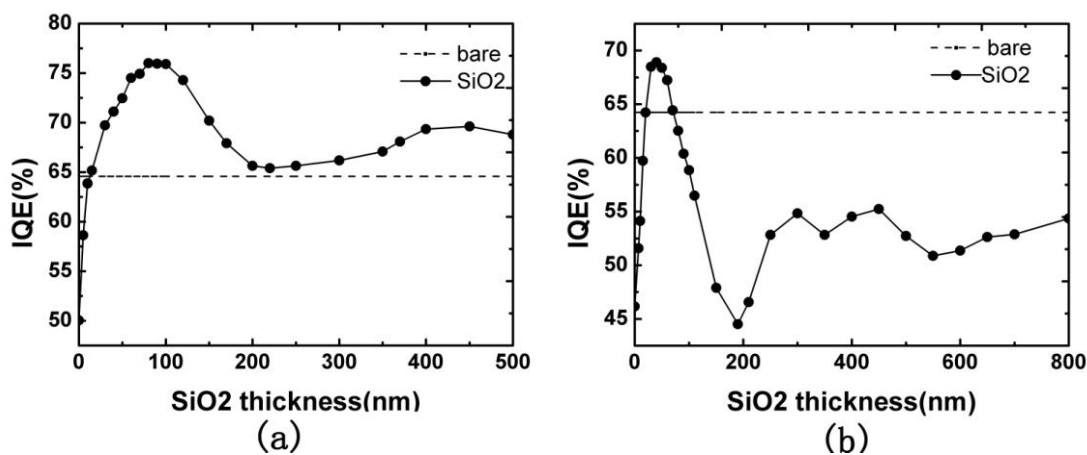

Fig. 5 The integrated quantum efficiency [$IQE$] vs. the thickness of $SiO_2$, for two different sizes: (a) radius=30 nm and (b) radius=70 nm.

Figure 5 shows oxide-thickness dependent $IQE$ calculated for Si substrates. Two pictures show a similar profile. Generally, when the dielectric layer is very thin, the absorption increases obviously with the spacer thickness. The absorption enhancement achieves the best performance as the thickness is less than 100 nm, e.g. 90 nm for 30 nm radius and 45 nm for 70 nm radius. The $IQE$ increases slowly and gets the maximum finally, as the dielectric layer is thick enough. Due to the influence of geometric optics, the oscillations are observed. Therefore,

SiO$_2$ layer of a suitable thickness is needed to achieve the best absorption. In addition, as can be seen from the above results, the nanoparticle size also plays an important role in determining the absorption performance, established obviously by the strong size effect on $Q_{scat}$, *CPE*, and *IQE*.

## 2.2. Varying the radius of the semisphere

Next the scattering behavior of semispheres with different sizes will be discussed, which are located on a 30 nm thick SiO$_2$, underlain by the Si substrate. The XY plane of the simulation domain is a W by W square, where W is the sum of radius of the semisphere and 40 nm. The boundary conditions are the same as the above simulation.

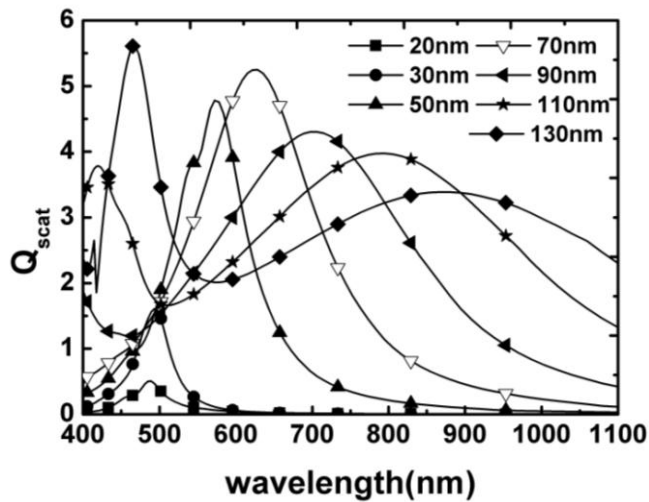

Fig. 6 The normalized cross section of Ag nanoparticles with different radii as a function of wavelength, illuminated by a total field scattered field source with a wavelength range from 400 nm to 1100 nm. The numbers at the top right corner indicate the radii of the Ag semishperes.

Figure 6 shows the normalized scattering cross section of metallic semispheres with different radii. According to the analytical calculation from Mie theory, the plasmon resonance is affected by a redshift as the radius is progressively increased. Figure 6 also shows the peak value of $Q_{scat}$ increases with the size as the radius is less than 70 nm. At resonance the scattering cross section can well exceed the geometrical cross section of the particle. For example, at resonance the metallic semisphere with 70 nm radius has a scattering cross section that is around 5 times of the geometrical cross section. In such case, to first-order, a substrate covered with a 20% areal density of particles could fully scatter the incident light. For the size well below the incident light wavelength, a point dipole model describes the scattering of light well. While an increased size leads to a larger absolute scattering cross section when the radius is of the same order as the light wavelength, and this modification leads to a reduced cross section when normalized by size. A narrower peak at short wavelength due to high-order excitation is also observed. For these large sizes, dynamic depolarization and radiation damping become important. Furthermore, the excitation higher-order plasmon modes must be taken into account for the large size [22-23,25]. Dynamic depolarization occurs as the particle size increases, because conduction electrons across the particle no longer move in phase. This leads to a reduction in the depolarization field at the centre of the particle [28]. As a result, there is a reduced restoring force and hence a redshift in the particle resonance. For these large particles, where scattering is significant, this re-radiation leads to a radiative damping correction to the quasi-static polarizability, inducing plasmon resonance broadening [29]. The redshift and broadening of the resonance would generally be expected to enhance optical absorption, since light-trapping should occur over a relatively broad wavelength range and more light at the band-gap wavelengths will be scattered into the Si substrate.

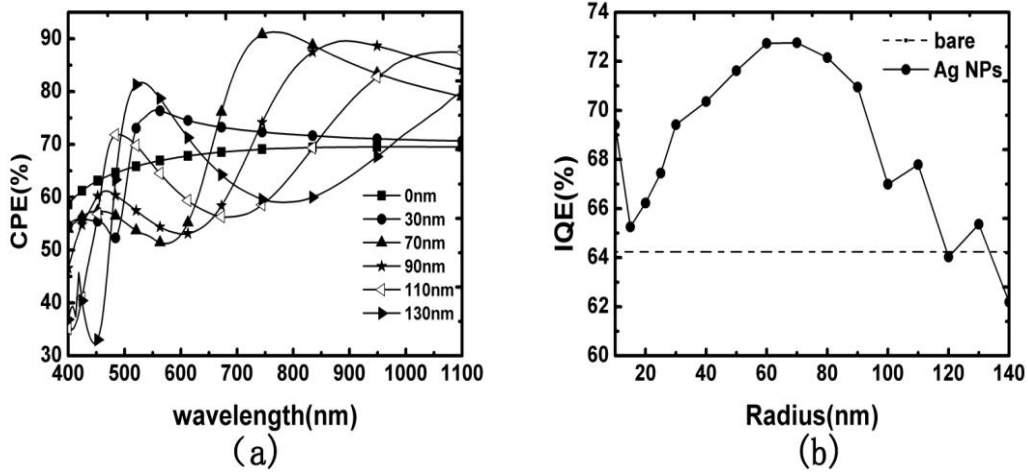

Fig.7 (a) The light coupling efficiency into the Si substrate and (b) the integrated quantum efficiency of different sizes

Figure 7 shows the fraction of light directed into the Si substrate and the integrated quantum efficiency for the particles with different sizes. Because of the higher-order modes, an evident peak is observed at the short wavelengths. As can be seen in figure 7, the coupling efficiency at high-order resonance is smaller than that of dipole modes, although the high-order modes show a great scattering cross section. In fact, high-order plasmon modes reduce the light coupled into a dielectric substrate because the field around the particle decays much faster than the dipolar resonance. The dipolar resonance mode and high-order plasmon modes determine the total optical coupling together, inducing some oscillations on the $IQE$ curves with the increase of the radius. In summary, an appropriate granular size should also be selected to achieve the best absorption performance for a certain thickness of the dielectric layer.

## 3. Conclusions

In this paper, we have investigated the spacer-thickness dependent metallic plasmon effect on silicon solar cells. As the oxide is very thin, the metallic plasmons depend on the Si substrate and the $SiO_2$. As the effect of the oxide on plasmons is the same order as that of the Si substrate, the resonance wavelength and $Q_{scat}$ peak value are characterized by oscillations. As the spacer is thick enough, the effect of the Si substrate on the metallic plasmons is negligibly small. Moreover, the resonance wavelength oscillates with the oxide thickness under the influence of the geometrical optics as the thickness is comparable to the light wavelength. The influence of the particle diameter has also been analyzed while keeping the spacer thickness constant. With the increase of particle size, the resonance peak redshifts and the normalized cross section increases. As the nanoparticle is large enough (usually the same order as the incident light wavelength), high-order excitation occurs, inducing a normalized cross section peak in the short wavelength range. Moreover, radiative damping induces the broadening of the plasmon resonance for the large particles. In conclusion, the optimization of plasmonic light trapping in a solar cell is a tradeoff, where the size of particle and the thickness of dielectric layer must be taken into account.